# Non-Kolmogorov Turbulence and Inverse Energy Cascade in Intracranial Aneurysm: Near-Wall Scales Suggest Mechanobiological Relevance

Simon Tupin[1], Khalid M. Saqr[1‡], Sherif Rashad[2,3], Kuniyasu Niizuma[2,3,4], Makoto Ohta[1], Teiji Tominaga[3]

(1) Biomedical Flow Dynamics Laboratory, Institute of Fluid Science, Tohoku University, Sendai, 980-8577, Miyagi, JAPAN.

(2) Department of Neurosurgical Engineering and Translational Neuroscience, Tohoku University Graduate School of Medicine, Sendai, Miyagi, 980-8574, JAPAN.

(3) Department of Neurosurgery, Tohoku University Graduate School of Medicine, Sendai, Miyagi, 980-8574, JAPAN.

(4) Department of Neurosurgical Engineering and Translational Neuroscience, Graduate School of Biomedical Engineering, Tohoku University, Sendai, Japan.

‡Corresponding author: Khalid M. Saqr, Ph.D.

Email: k.saqr@tohoku.ac.jp , kh.saqr@gmail.com

## ABSTRACT

The genesis, growth and rupture of intracranial aneurysm (IA) are open questions in neurovascular medicine until the present moment. The complexity of aneurysm mechanobiology and pathobiology staggeringly combine intertwining biological and physical processes that are tightly connected. Recently, transition to turbulence in IA blood flow is thought to play a central role in IA growth and rupture as it can be directly linked to endothelial dysfunction. However, the problem of turbulence brings unprecedented complications to the topic. We found turbulence in IA to be of non-Kolmogorov type. For the first time, we detected inverse kinetic energy cascade in blood flow and in non-Kolmogorov turbulence. Here, we hypothesize that the near-wall turbulence undergoing inverse energy cascade have scales that could affect the mechano-signaling of endothelial cells. Our findings could be a paradigm shift in the contemporary theory of aneurysm hemodynamics.

**BACKGROUND**

Rupture of intracranial aneurysm (IA) is a major cause of nontraumatic subarachnoid hemorrhage (SAH). Aneurysm pathophysiology and mechanobiology are directly influenced by hemodynamics [1]. Experimental measurements[2] and numerical simulations[3-5] showed that blood flow in intracranial aneurysm (IA) exhibits transition to turbulence[6] and direct energy cascade[7]. Disturbed flow (i.e. transitional/turbulent) promotes proinflammatory and degenerative pathways in endothelial cells (ECs)[8]. Nevertheless, the exact flow parameters which influence different biological pathways and mechanisms remain unclear[9]. Computational Fluid Dynamics (CFD) contributed extensively to the hemodynamic theory of IA in the past three decades, however, resulted in numerous controversies and ambiguities. A complete account of the role of hemodynamics in IA pathobiology can be found in the recent review and meta-analysis by Saqr et al [10]. Some computational studies in the past few years pointed the importance of considering transition to turbulence in IA as a possible hemodynamic variable that could influence its growth and rupture [11-13]. However, there is no complete theory that correlates the features of IA transitional/turbulent flow to its pathophysiology and mechanobiology. In simple words, we know that the flow in IA is transitional/turbulent, however we do not know the characteristics of such flow and we do not know how such flow affects IA on cell and tissue levels. The present work sheds the light on these questions via experimental measurements.

Fluid turbulence is one of the everlasting mysteries of classical mechanics. Therefore, it is important to briefly revisit the theory of turbulence to appreciate the findings of the present work. Turbulence is chaotic oscillations of a flow field that result in vortex cascade phenomena. Richardson conceptualized the phenomena of vortex cascade in turbulent flows[14]. Kolomogorov formulated a statistical theory that describes the cascade of turbulence kinetic energy in homogenous isotropic turbulence[15]. Such theory dictates that kinetic energy transfers from large slow vortices to smaller and faster ones[16]. Then, two decade later, Kraichnan predicted the possibility of inverse energy cascade in two-dimensional Kolmogorov turbulence [17]. Kraichnan predictions were validated in later experimental [18, 19] and numerical [20] works and became an essential part of the statistical theory of turbulence. Then, inverse energy cascade, associated with isotropic homogenous turbulence, has been detected in geophysical flows [21-23] and most recently in Jupiter's atmosphere[24]. However, until the present moment there is no evidence on the existence of inverse energy cascade in non-Kolmogorov turbulence nor in biological flows.

Despite the fact that transition to turbulence of simple monoharmonic pulsatile flow in straight pipes is not fully understood [25, 26], substantial evidence show that such flow could undergo transition to turbulence at $Re_m = \frac{\rho u_m D}{\mu} < 1000$ and $\alpha = R\sqrt{\frac{\omega \rho}{\mu}} < 3$ [27-29]. Blood flow is a multi-harmonic nonsynchronous oscillating flow with 95% of the frequency spectrum falling below 12 Hz[30]. In addition to blood's waveform complexity, anatomical variations of arteries and aneurysm geometries dictates shear-driven vortex formation, shedding, precessing, and a multitude of complex flow physics that are yet to be unveiled[26, 31]. Direct numerical simulations of intracranial aneurysm hemodynamics showed that an intermittent direct energy cascade exists at $Re_m \sim 500$ and $\alpha \sim 4$ [7, 32]. The primary objective of this article is to prove the existence of non-Kolmogorov turbulence in pulsatile flow in an IA model at Reynolds number less than 400 and Womersley number less than 3. The secondary objectives are to characterize some of the features of such turbulence and propose a hypothesis to combine such features with the recent findings of endothelial cell response to local forces applied on the cell cytoskeleton. We have used well established methodology to conduct the Particle Image Velocimetry (PIV) measurements and analyze the results in frequency domain.

**MATERIALS METHODS**

An ideal ICA sidewall aneurysm geometry was fabricated as shown in figure 1. It was composed of a cylindrical artery of diameter d = 4 mm and a sphere of diameter D = 10 mm located at a distance of 6.25 mm [33, 34]. This geometry was used to mould $200 \times 50 \times 50$ mm³ box-type models made of silicone (R'Tech, Japan), considered as rigid, and 12 wt.% PVA-H (polyvinyl alcohol hydrogel), for greater compliance [35], following a manufacturing process previously described in detail [36, 37]. A pulsatile flow pump (Alpha Flow EC-1, Fuyo Corporation, Japan) was used to reproduce physiological flow conditions, as shown in figure 2, with a flow rate ranging between 200 and 300 mL/min [38], and pressure between 70 and 120 mmHg [39], with waveforms shown in figure (1-b). The waveform period was $T$ = 1 s and the time to peak systole was 0.3 s. Coriolis flow-meters (FD-SS2A, Keyence, Japan) and pressure sensors (AP-12S, Keyence, Japan) were used to monitor instantaneous flow conditions at the inlet and outlet of the models. Fourier decomposition of the inlet flow rate waveform revealed the complex physiological flow reproduction composed of 9 harmonics (figure 1-c). A blood-mimicking fluid, made of a mixture of glycerin (42.8%), water (47.6%) and sodium iodide (9.6%), was used to approximate blood kinematic viscosity [40]($v_f = 4.0 \times 10^{-6}$ m²/s) and density [41] ($\rho_f$= 1270 Kg/m³), and match models refractive index [42] ($n_f$ = 1.41). The oscillating mean Reynolds and Womersley numbers were calculated as following: $Re_{os} = \frac{\rho \tilde{u} d}{\mu}, Re_m = \frac{\rho U_m d}{\mu}, \alpha = r\sqrt{\frac{\omega \rho}{\mu}}$. The inflow waveform used in our experiment was multiharmonic, constituted of nine harmonics of cardiac frequency spectra ($f_1 = 1\,Hz, 2 \leq f_n \leq 9\,Hz$) superimposed to the mean blood flow (figure 1-b and 1-c). The flow oscillating, mean Reynolds and Womerlsey numbers were $271.89 \leq Re_{os} \leq 397.88, Re_m = 313.45, \alpha = 2.51$, respictively. The experimental setup is schematically shown in figure 2. The Fourier coefficients of the Womerlsey flow are detailed in table 1.

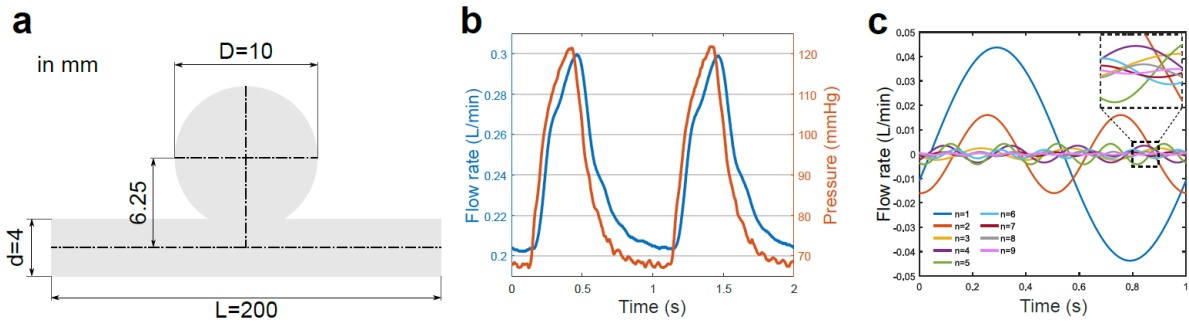

**Figure 1. Aneurysm model geometry and inflow waveform with its Fourier decomposition.** (a) Schematic of the aneurysm model geometry and dimensions, (b) the inflow and pressure waveforms as measured at the inlet of the model, the phase difference between flow and pressure waves is a well-documented property of the Womersley solution of Navier-Stokes equation[1] and is attributed to the flow inertia which is proportional to the inflow waveform frequency. (c) Fourier decomposition of the inflow harmonics showing the nine main oscillating components and the inset shows the complexity of the secondary harmonics ($n = 2:9$) which results in the self-sustained oscillations.

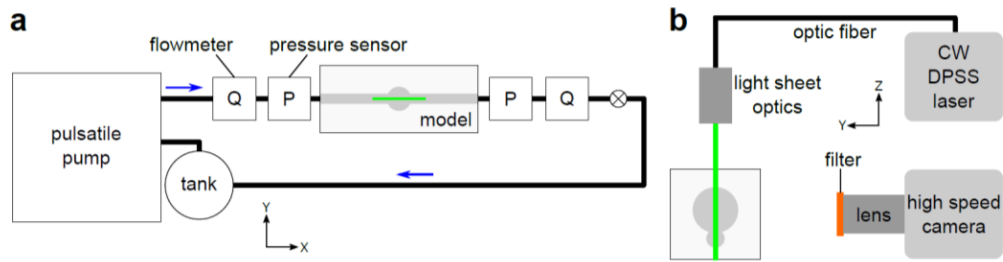

**Figure 2. Schematic of the experimental and laser PIV measurement setup.** (a) the flow circuit used in the experiment consists of a closed loop comprising a pulsatile flow pump generates physiological flow, surge tank, upstream and downstream flowmeters and pressure transducers connected by fixed 4 mm inner-diameter tubes and the pressure is controlled by a check valve

downstream the aneurysm model. The tubes used were all rigid and the measurements were filtered to remove any system generated noise. (b) the laser PIV measurement setup showing the measurement plane and high-speed camera arrangement.

**Table 1.** Values of the Fourier decomposition coefficients and phase shifts of the physiological flow signal. The decomposition is given by $Q(t) = q_o + \sum_{n=1}^{8} a_n cos(n\omega t - \phi_n)$. The instantaneous spatially averaged flow rate $Q(t) = \frac{\pi}{4} d^2 u_{ave}(t)$, where $u_{ave}(t) = \frac{1}{2}\left(u_o\left(1 - \frac{r^2}{R^2}\right) + \sum_{n=1}^{8} b_n cos(n\omega t - \phi_n)\right)$ where $u_o$ is the steady centerline velocity component and $b_n$ is the magnitude of each velocity harmonic where $a_n \neq b_n$.

| n | a | phi |
|---|---|---|
| $q_o$ | 0.236341 | |
| 1 | 0.043718 | 2.804445 |
| 2 | 0.016012 | -1.11386 |
| 3 | 0.002393 | 1.02666 |
| 4 | 0.003461 | -0.01789 |
| 5 | 0.004199 | 2.304288 |
| 6 | 0.001759 | -1.57197 |
| 7 | 0.0008 | -1.98073 |
| 8 | 0.000964 | 0.447477 |
| 9 | 0.000313 | 2.747776 |

Rhodamine B encapsulated microspheres (FLUOSTAR, EBM, Japan) of 15 µm diameter were selected as fluorescent particles along with a long-pass filter (cut-off wavelength at 550 nm) placed in front of the camera. The PIV acquisition system was composed of a high-speed camera (Fastcam Mini UX100, Photron, Japan), mounted with a lens ($f$ = 105 mm), and a CW DPSS laser (Millennia eV, Spectra-Physics, USA) with a power of 2W and a wavelength of 532 nm. A light sheet optics module (80X91, Dantec Dynamics, Denmark) was used to generate a ≈ 1 mm thickness laser sheet to illuminate the center plane of the model, where out-of-plane velocity component is assumed negligible due to model symmetry [43]. A total of 12000 images (1024 × 1024 pixels, 19.7 µm by pixel) were captured at a frame rate of 2000 Hz and a shutter speed of $2\times10^{-4}$ s. Velocity fields were computed by cross-correlation of consecutives images (DaVis 8.4.0, LaVision GmbH, Germany). An adaptive multi-pass scheme (interrogation windows refined from 96 × 96 px to 48 × 48 px) was adopted with 75% overlapping and adaptive PIV option for the last pass. This adaptive PIV approach, as previously described [44], calculates the optimal local interrogation window size and shape based on flow gradients and image quality. The interrogation window size is varied according to the flow gradient (standard deviation of the displacement field σ inside the interrogation window) and the correlation value, C, using a local weighting factor, W, defined as: W = 1/(C · σ)

The shape of the window is set as an elongated Gaussian weighted ellipse with principal direction, α, and eccentricity, ε, computed by minimizing the differences of displacements within the window $dV_\alpha = \sum_{\alpha=0}^{180} |v_0 - v_\alpha|$ and $\varepsilon = \frac{4}{1+3(dV_\alpha/dV_{\alpha+90})}$

Due to the lower flow velocity in the aneurysm compared to the vessel, analysis of the aneurysmal flow was performed using 1 image over 4 (equivalent to 500 Hz). The local kinetic energy in frequency domain was calculated as [45] $E_i(f) = \frac{|\mathcal{F}\{\tilde{u}_i(t)\}|^2}{L \cdot f}$ where f the frequency, F the fast Fourier transform and L the length of $\tilde{u}_i$ matrix. Enstrophy was calculated by integrating the vorticity ($\omega(t)$) across the aneurysm domain using cumulative trapezoidal numerical integration (CTNI) at every time step. Then, enstrophy cascade was obtained by Fourier transform of the resulting integration. The local dissipation length scales were calculated as $\frac{1}{\lambda_i^2} = \frac{2\pi^2}{\overline{\tilde{u}_i}^2 \sigma_{\tilde{u}_i}^2} \int_0^\infty f^2 E_i(f) df$ where $\overline{\tilde{u}_i}$ and $\sigma_{\tilde{u}_i}$ are the mean and standard deviation of $\tilde{u}_i$, respectively. This parameter was evaluated for each point in the domain to plot $E(f, x_i)$. A two-dimensional local smoothing function [46] was applied to improve the readability of the 3D surface plots and better visualize the qualitative changes of energy or vorticity versus length scale and frequency.

**RESULTS AND DISCUSSION**

Using physiologically relevant *in vitro* intracranial aneurysm model, we investigated the kinetic energy cascade in frequency domain without using Taylor's hypothesis. The inverse cascade was detected in multiharmonic pulsatile flow using ms/µm laser PIV in rigid (silicone) and elastic (PVA hydrogel) side-wall aneurysm models suggesting its independence from wall compliance. The vessel to aneurysm diameter ratio used in this study was $\frac{2}{5}$ (see figure 1-a for model dimensions) which corresponds to moderate to large intracranial aneurysm size. Random transitional/turbulent oscillations during six flow pulses are clearly shown in the aneurysm in comparison with the periodic pattern observed in the parent artery, shear layer and separation region, as shown in figure 3. Through a distance as small as 2 mm, between the vessel center and the shear layer (subfigures C and D in figure 3, respectively), the flow transits from periodic to quasi-periodic pattern, and the value $\left|\frac{\tilde{u}}{U_m}\right|$ increases one fold. In non-homogenous anisotropic regimes, where $\left|\frac{\tilde{u}}{U_m}\right| > 1$, the use of Taylor's frozen turbulence hypothesis is inappropriate[47, 48] and the energy cascade is only available in frequency domain $E(f)$ rather than the wavenumber domain $E(k)$. Also, we found that Kolomogorov's $-\frac{5}{3}$ cascade and corresponding scales do not apply (see figure 4 and table 2).

In table 2, the (+,-) signs denote inverse and direct energy cascades, respectively. Kolmogorov-Obukhov theory of 1941 and its substantiation of 1962 inherited the homogeneity and isotropy from Taylor's 1937 theory, and the former dictates that the rate of decay of kinetic energy as function of frequency or wavenumber (since wavenumber is linearly proportional to frequency in Taylor's hypothesis) should be homogenous in Cartesian and ensemble spaces and isotropic in directionality. This does not apply to the negative (direct) decay rate values shown in table 1. The Kraichnan inverse energy cascade theory, proposed in 1967, was based on the work of Kolmogorov, hence it only applies in homogenous isotropic turbulence. This also does not apply on the positive (inverse) energy cascade shown in table 1. It should be noted that the unavailability of wavenumber in this work only shows the inverse cascade as function of frequency which should intuitively have the same meaning of Kraichnan inverse cascade. The fact that energy cascade has different power values at different locations, where the flow exhibits random and high velocity/frequency oscillations, stipulates the non-Kolmogorov nature of the flow. In order to confirm our findings, we conducted similar measurements using PVA-H (polyvinyl alcohol hydrogel) aneurysm models. The inverse energy cascade was detected in both models which indicates that the existence of inverse energy cascade, as well as the non-Kolmogorov turbulence, are independent from wall compliance, as shown in figure 5.

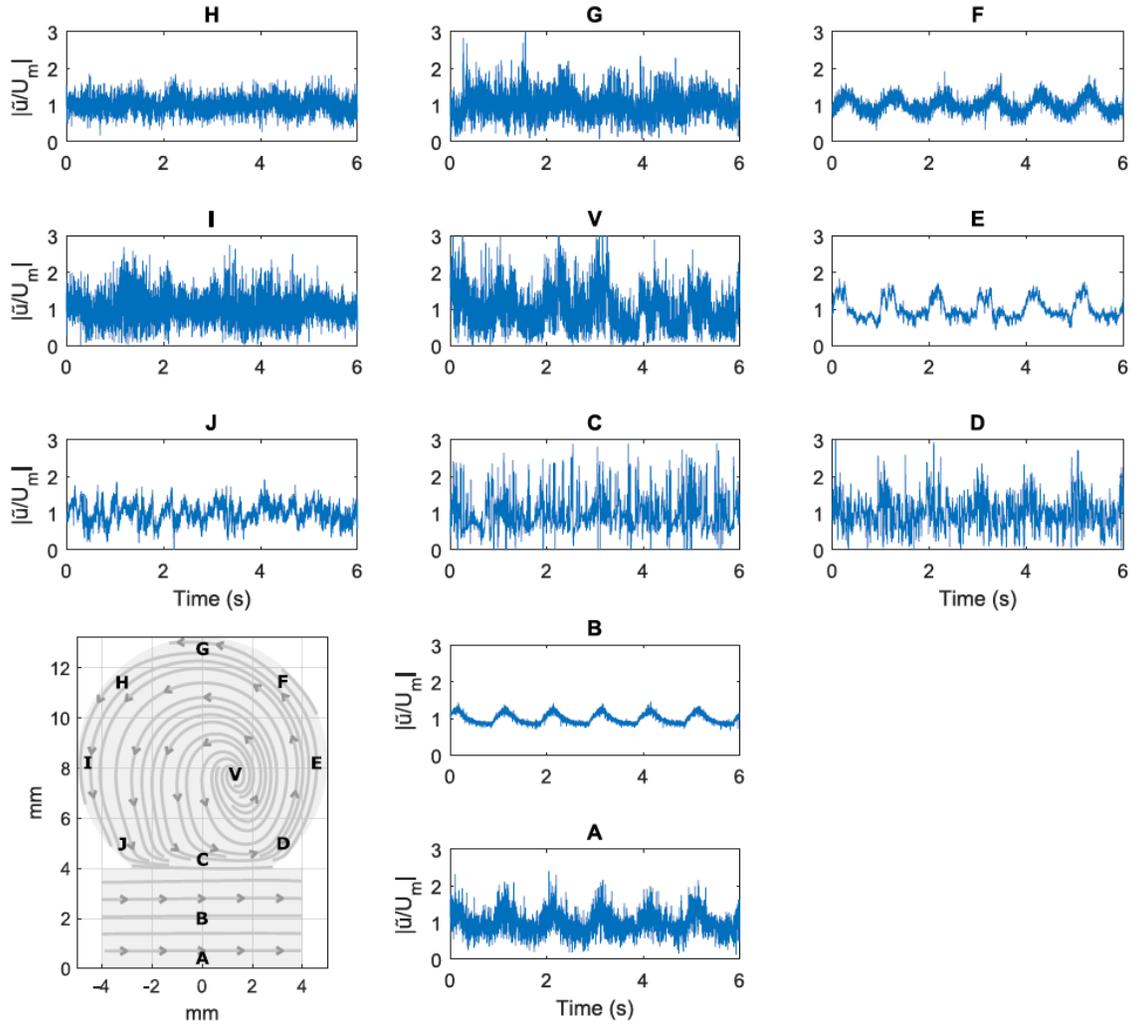

**Figure 3. Velocity history profiles at selected locations in the vessel and aneurysm showing local turbulence characteristics.** Six-seconds time series of oscillating velocity magnitude normalized by local mean velocity magnitude ($|\tilde{u}/U_m|$) at selected points near the aneurysm wall and in the vessel. The points are indicated on the aneurysm model by letters (A-J) in the lower left inset. The periodic behavior is conserved in the vessel (A,B) and in the entry jet region (E). In contrast, random velocity fluctuations are observed in the shear layer (C), in the separation region (D) and near the aneurysm wall (F-J). The random velocity fluctuations are obvious marking local (i.e. non-homogenous) transitional features in the aneurysm. The values of $|\tilde{u}/U_m| > 1$ and its spatial variation within the aneurysm geometry (D=10 mm) indicates large (i.e. anisotropic) structures.

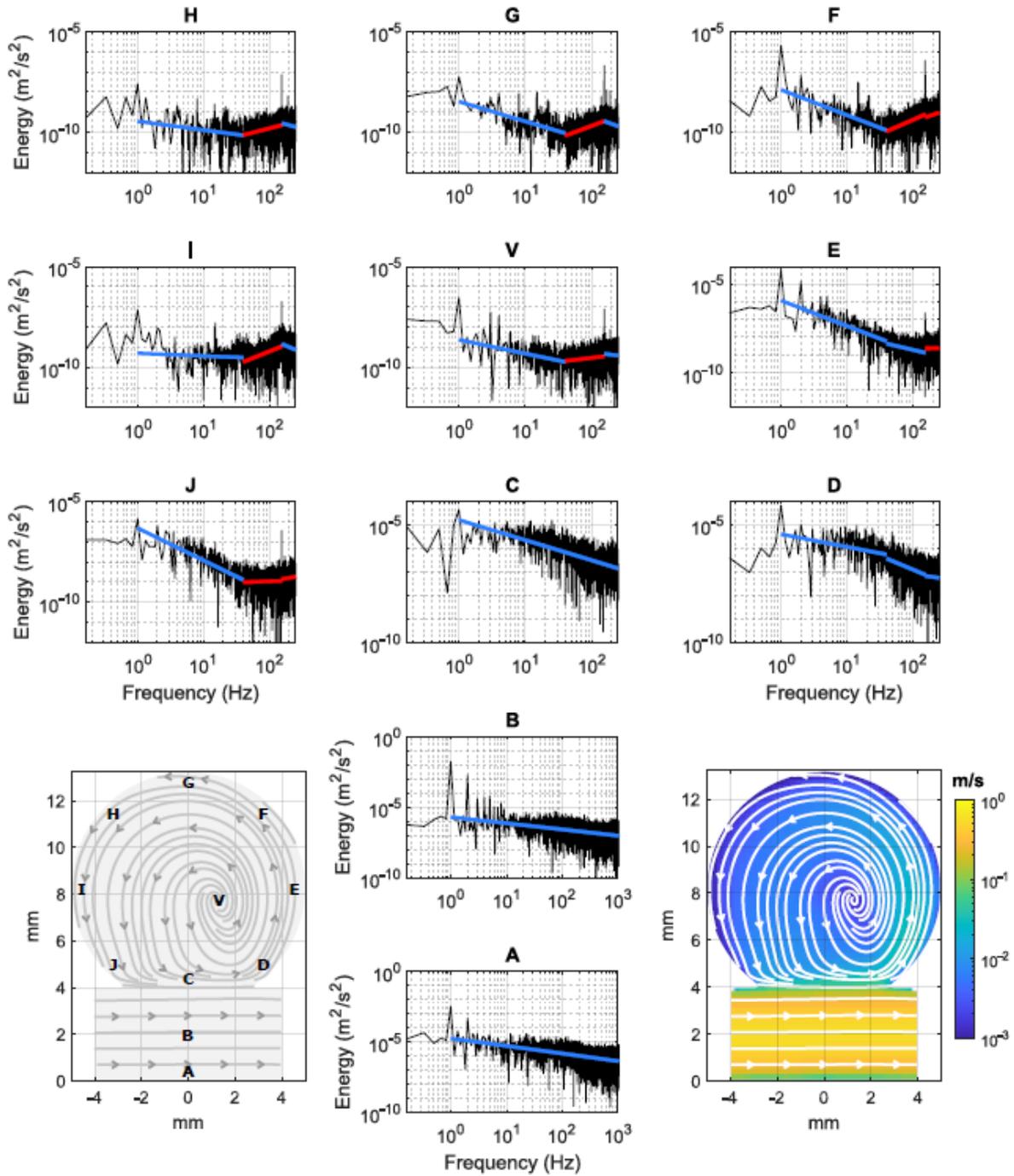

**Figure 4. Inverse energy cascade is detected near the aneurysm model wall and not in the vessel**. Various points taken at different near-wall locations are shown on the lower left inset, while the absolute time-averaged velocity and streamlines are shown on the lower right inset.

**Table 2. Energy cascade power values show variation that indicate non-Kolmogorov cascade and scales.** The cascade power values, approximated to 3 decimal figures, at the selected points shown in fig. 1.

| point      | A      | B      | C      | D      | E      | F      | G      | H      | I      | J      | V      |
|------------|--------|--------|--------|--------|--------|--------|--------|--------|--------|--------|--------|
| [1:end]    | -0.530 | -0.434 | -0.855 |        |        |        |        |        |        |        |        |
| [1:40]     |        |        |        | -0.539 | -1.440 | -1.253 | -0.977 | -0.436 | -0.133 | -1.599 | -0.684 |
| [40:154]   |        |        |        | -1.17  | -0.868 | 1.431  | 1.230  | 0.829  | 1.312  | 0.127  | 0.385  |
| [154:end]  |        |        |        | -0.383 | 0.029  | 1.047  | -1.149 | -0.847 | -1.381 | 0.549  | -0.377 |

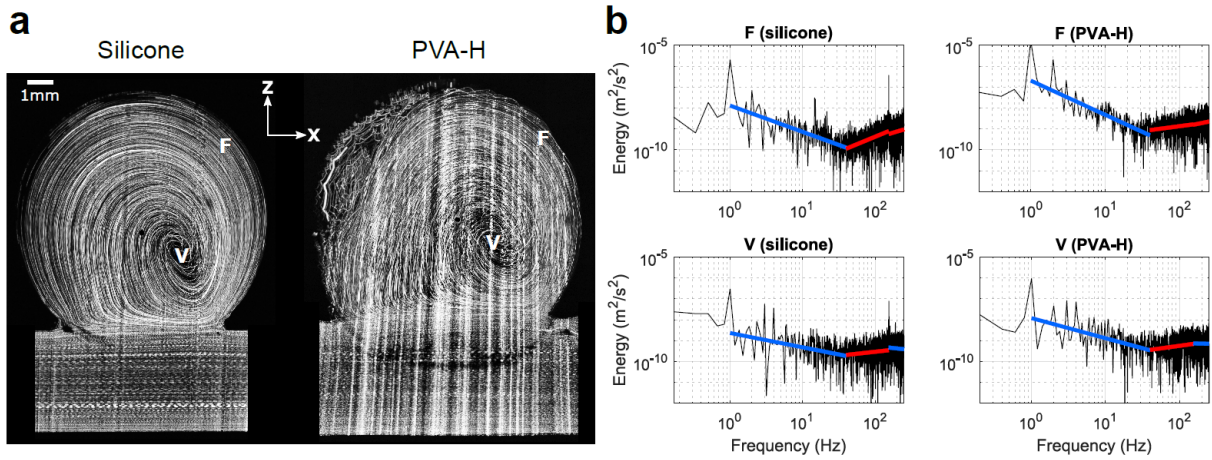

**Figure 5. Inverse energy cascade is independent from wall compliance as indicated by comparing rigid and elastic models.** (a) Photographs of time-averaged traces of fluorescent particles in silicone (rigid) and PVA-H (elastic) models. The photograph of the elastic model is less clear due to the elastic model compliance. (b) examples of energy cascades at the vortex core (V) and at the first point downstream the entry jet region (F) showing the inverse cascade behavior in both models.

In figure 4, we show $E(f)$ portraits at selected points in the parent vessel and near aneurysm wall, where the ECs would be subjected to pathologic flow predisposing to dysfunction and/or rupture. The energy cascade in the main vessel, where the flow is apparently periodic (figure 3, sub B) is shown to be direct with no major energy transfer events at frequencies higher than 10 Hz (see figure 4, sub A and B). Near the aneurysm wall (figure 3, sub E to J) and in the aneurysm vortex core (figure 4, sub V), the flow exhibits inverse energy cascade. To observe the inverse cascade in space domain, one-dimensional spatiospectral visualization of the reconstructed $E(f)$ field on representative lines (figure 6-a) depicted the inverse energy cascade around the vortex core (figure 6-b) and in the flow recovery region (figure 6-c) and not in the vessel (figure 6-d). Near the wall, where inverse energy cascade was observed, energy levels at $f \geq 200\ Hz$ were of similar values to the energy levels of the dominent harmonics $1 \leq f \leq 10\ Hz$. In the main vessel, however, the energy cascade was only direct as shown in fig. 2D. The values of energy in the inverse cascade regions had the order of $0.1\ \mu J/kg$.

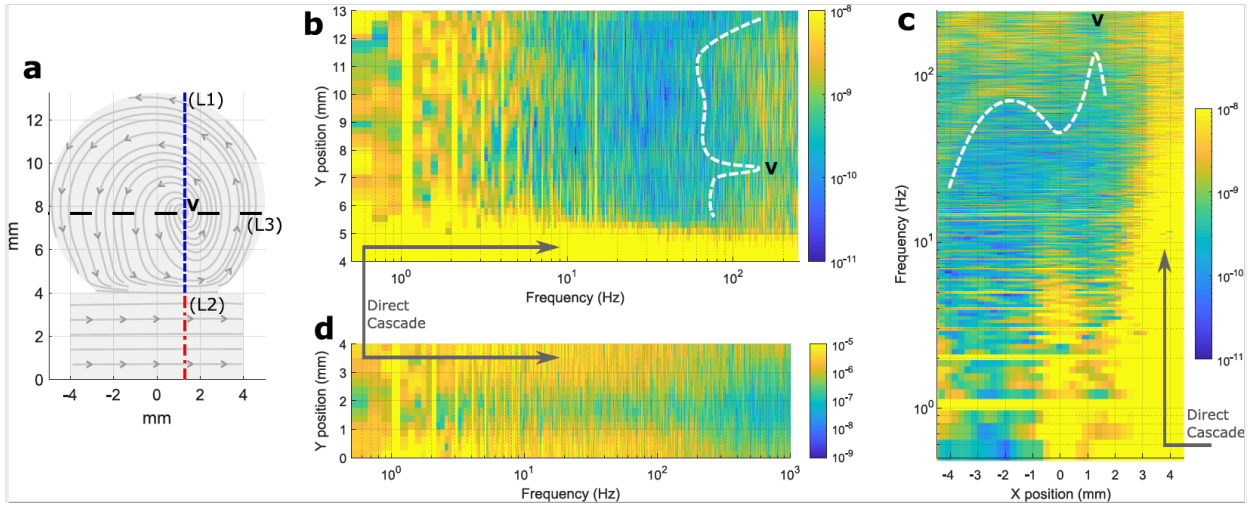

**Figure 6. Inverse energy cascade occur at high frequency spectra around the vortex core and near the aneurysm wall.** One dimensional spatiospectral $E(f)$ maps on vertical and horizontal lines passing across the aneurysm vortex core as depicted in (a). The direction of direct cascade in terms of frequency spectrum is indicated by arrows on the vertical lines (L1) and (L2) passing across (b) aneurysm and (c) vessel, respectively. The locations where inverse energy cascade is observed are denoted by dotted white lines in (b) and on the horizontal line (L3) passing across the vortex core denoted by (V). Inverse cascade takes place around the vortex core (b) and in the vortex recovery region (d) for frequency range $f \geq 80\ Hz$ and $f \geq 50\ Hz$, respitively.

In order to visualize length scale $\ell_s$ without using Taylor's hypothesis; time averaged field of $\ell_s$ was calculated locally at each measurement point and plotted as color map in figure 7-a (see methods section for details). We found that $\ell_s$ near the aneurysm wall had the order of $1 \sim 10\ \mu m$ (figure 7-a). In generalized two-dimensional Cartesian space, the kinetic energy in the vessel and aneurysm were plotted against length scale and frequency in figure 7-b and 7-c, respectively. We identified inverse energy cascade events at $f > 100\ Hz$ corresponding with $\ell_s = 1 \sim 10\ \mu m$ (fig.

7-d). The energy levels at such values of $\ell_s$ have the order of $10^{-3} \sim 10^{-4} \mu J/kg$. As we observed the inverse energy cascade, we noticed that at certain frequencies the inverse cascade spans all the aneurysm domain, therefore, we selected some of these frequencies to study the energy as function of the space domain, as in figure 8. At $f = 154.33\ Hz$, as shown in figure 8-a, the aneurysm domain exhibits a spike of kinetic energy which brings the level of energy to similar levels of the dominant frequency($f = 1\ Hz$). This energy spike is thought to be a hydrodynamic resonance resulting from self-sustained oscillations arising from the multiharmonic flow. Similar self-sustained oscillations have been briefly documented in the literature[49], however, this is the first observation in multiharmonic pulsatile flow.

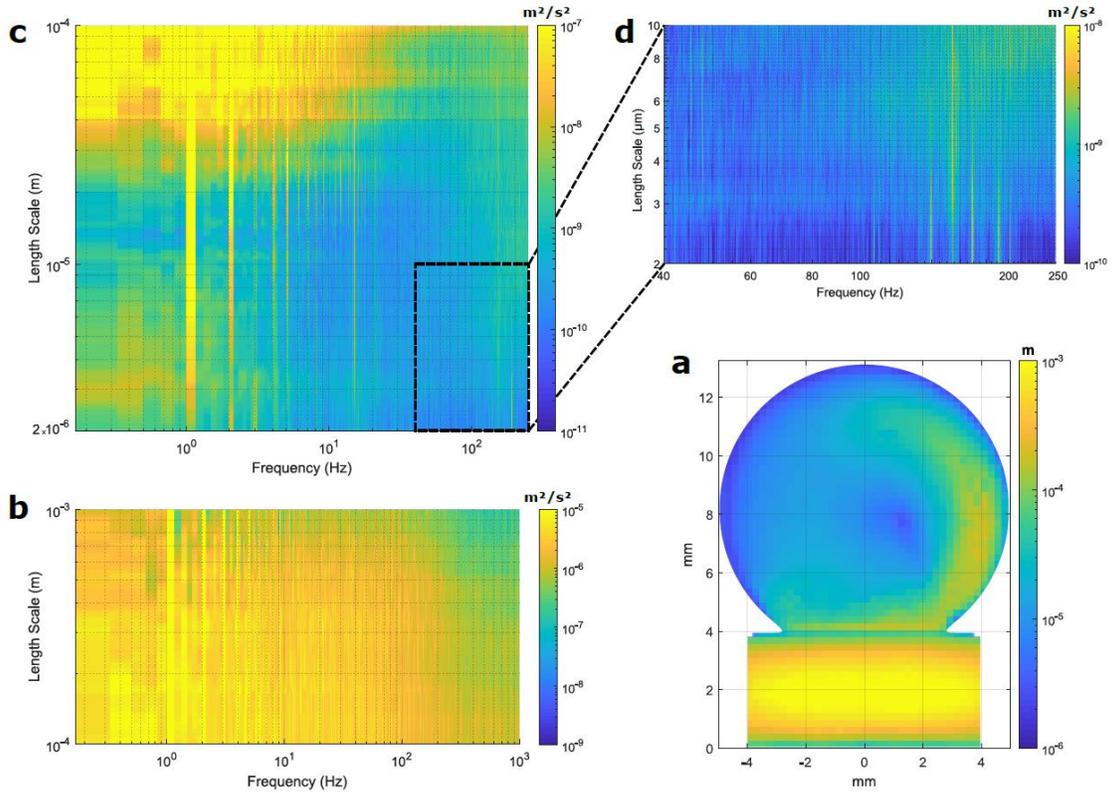

**Figure 7. Near the aneurysm wall, dissipative length scales are $1 \sim 10\ \mu m$ in size with kinetic energy level of $10^{-3} \sim 10^{-4} \mu J/kg$.** Time-averaged dissipative length scales are calculated locally at each measurement point and plotted in color maps in (a). Two-dimensional spatiospectral maps of $E(f)$ as a function of the time-averaged dissipative length scales in the vessel and aneurysm are shown in (b) and (c). The dominant (inflow) harmonics are depicted with the maximum energy shown as bright yellow lines in (b) vessel and (c) aneurysm at $1 \leq f \leq 12$. The inset (d) shows the details of the inverse cascade in the aneurysm at $40 \leq f \leq 250\ Hz$.

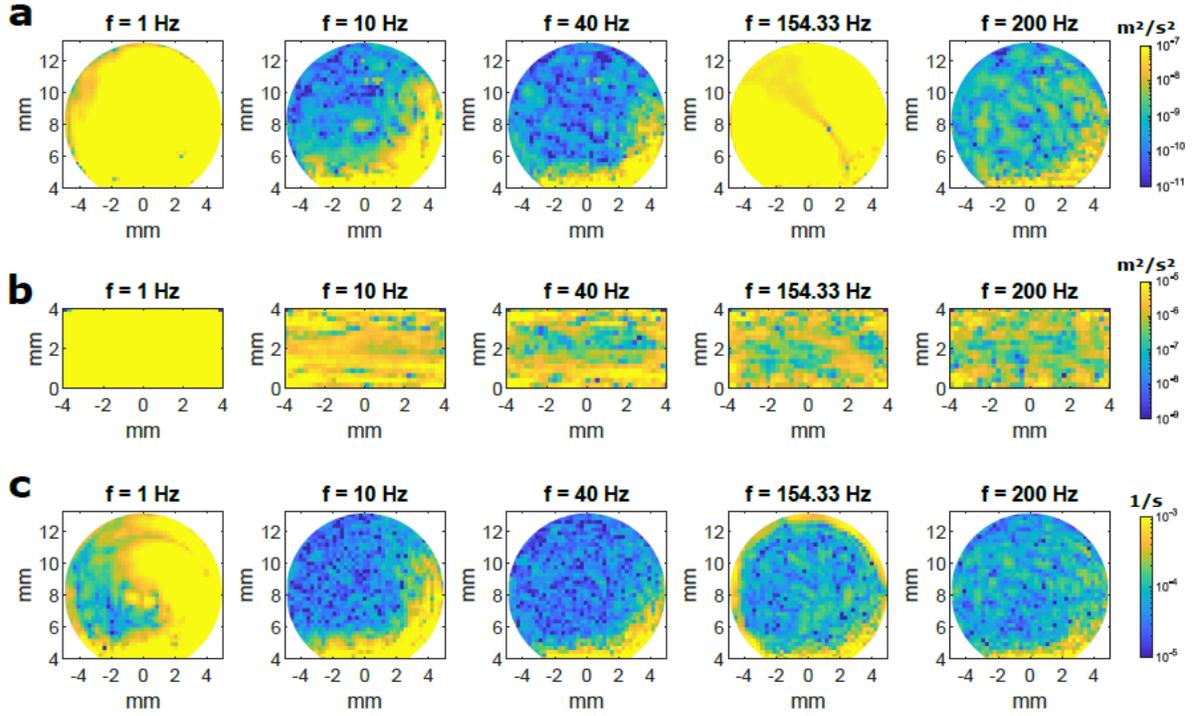

**Figure 8. Kinetic energy in the aneurysm (a) and vessel (b), as well as the local vorticity field (c) as function of location in mm at selected frequencies indicate resonance at $f = 154.33\ Hz$.** The direct cascade is evident in the vessel from 1 Hz to 200 Hz. In the aneurysm, a critical frequency of 154.33 Hz was observed where the energy budget increases significantly in the flow domain which denotes transition between direct and inverse cascades, and shows the resonance mode arising from the self-sustained oscillations.

In an attempt to further investigate such phenomena, local vorticity in frequency domain was investigated, as depicted in figure 8-c, and a non-characteristic near-wall high vorticity ring was detected at the critical frequency. In correlation with the length scales, the sudden increase in vorticity magnitude at the critical frequency was correlated with length scales as low as $2\ \mu m$, see figure 9. Enstrophy ($\epsilon_\omega = \int_S \omega^2(t)\ ds$) was calculated and plotted in frequency domain, see figure 9-c. Enstrophy is a global indication for the energy cascade, and the depiction of power-change of its cascade at $f \geq 40\ Hz$ confirms the existence of inverse cascade. Dissipative length scales as small as $\ell_s = 1\sim10\ \mu m$ exhibit inverse cascade at high frequency, and with energy content in the order of $0.1\sim1\ nJ/kg\ (10^{-9}\sim10^{-10} J/kg)$. With the former length scale, such energy level corresponds to instantaneous force of the order of $10^{-3}\ pN$.

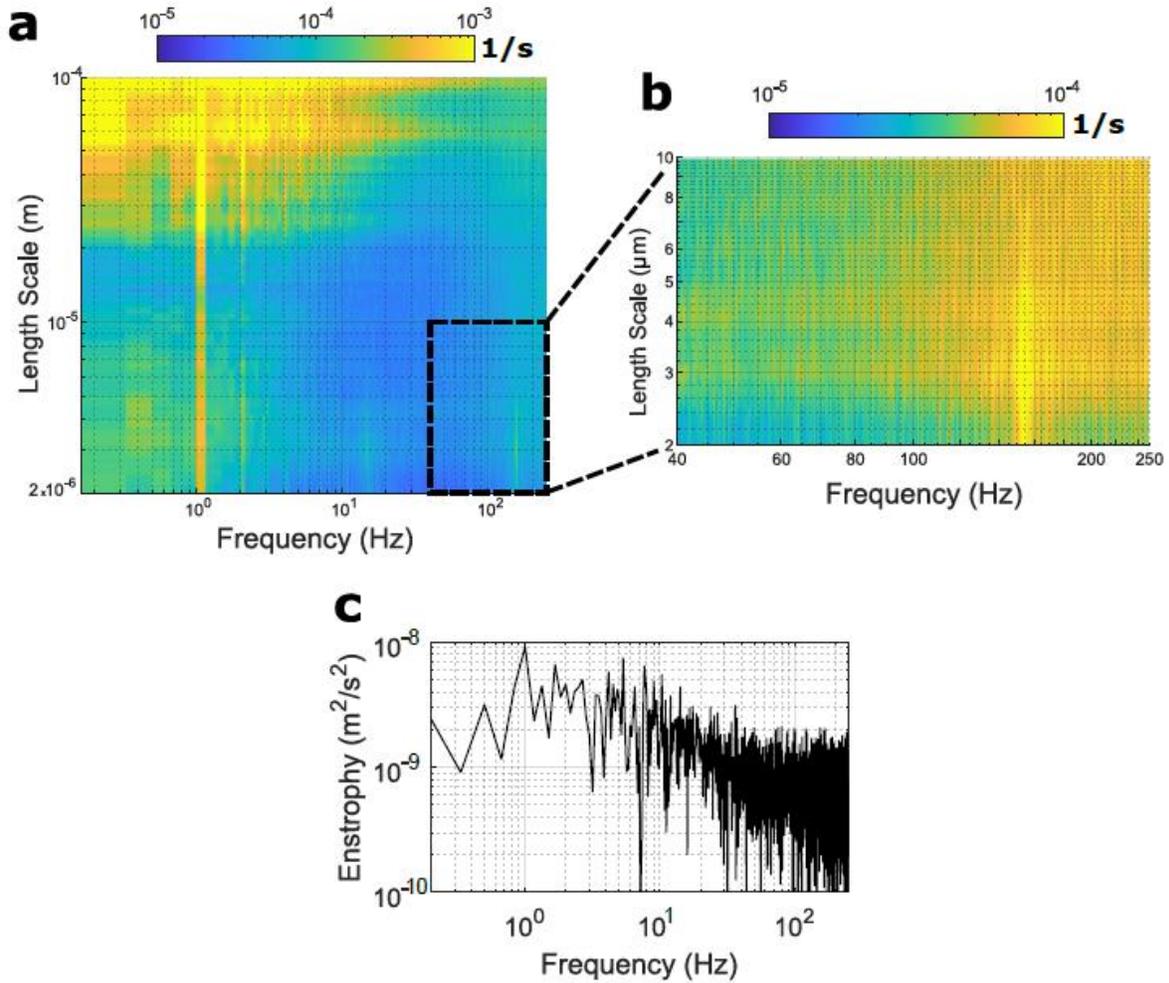

**Figure 9. Enstrophy cascade and vorticity correlation with length scale and frequency confirms the existence of inverse energy cascade.** The vorticity field in frequency domain is correlated with dissipative length scales in (a) and the spectra where the critical frequency denoting self-sustained oscillations are shown in inset (b). Global enstrophy field ($\epsilon_\omega$) is traced in frequency domain (c) and the change of its cascade at ($f \geq 40\ Hz$) marks the inverse cascade.

ECs sense flow via a variety of cell surface and intracellular mechano-receptors[50]. One of the main mechano-transducers is the cell membrane glycocalyx (GCX)[51]. Glycocalyx was shown to sense and transduce the flow induced forces into cellular response [52]. GCX also was shown to undergo structural organization in response to flow, and these structural changes can impact the endothelial cellular response [51]. When we analyzed the near-wall $\ell_s$ that demonstrated the inverse energy cascade phenomenon, they fell into the micrometer range ($1\sim10~\mu m$) Interestingly, it was recently shown that focal tension forces, in pN order of magnitude and at the micrometre length scale, applied via membrane tethers, are capable of activating the Piezo channels, inducing $Ca^{2+}$ entry and downstream cellular signaling cascade[53]. Thus theoretically, kinetic energy transfer at such scale may induce ECs' signaling cascade and a corresponding aberration may induce pathogenesis. ECs within aneurysms are known to have a pathologic phenotype, however the actual aneurysmal flow is still scrutinized and under debate [54]. While this holds true for aneurysmal ECs, the ECs within the parent vessel are exposed to physiologic flow stress and thus exhibit a physiologic phenotype [51]. Thus, we argue that the exposure to an inverse energy cascade in the aneurysm may induce a pathologic phenotype, which may explain the mechanobiological response to the aberrant flow field. Therefore, the exact effect of the inverse energy cascade on ECs should be investigated, and the effect of different energy cascades should be characterized. This will eventually lead to substantial improvement of the current hypothesis [1] about the mechanobiological role of wall shear stress (WSS) by including the role of local energy cascade and dissipation.

## CONCLUSION

Blood flow in intracranial aneurysm appears to be much more complex than previously thought. The observation of non-Kolmogorov turbulence and inverse energy cascade in such flow constitute staggering shift in the aneurysm hemodynamics paradigm. Near-wall turbulence possesses length and force scales that could be significant and influential in the ECs mechano-signaling. On the other hand, from the viewpoint of turbulence physics, it is indeed overwhelming to observe such complex phenomena at $Re < 400$ and $\alpha < 3$. The complex dynamics of multiharmonic flow as well as the inherited geometrical sophistication of aneurysms trigger series of physical phenomena that are yet to be discovered. The present work points to the fact that the classical differentiation between laminar and turbulent flow regimes, based on the Reynolds criteria for pipe flow, is inappropriate in the aneurysm problem.

## ACKNOWLEDGMENTS

The authors thank Mr. M. Matsuura for experimental assistance.

**Funding:** This work was supported by JSPS KAKENHI grant (18K18356), ImPACT program of Council for Science, Technology and Innovation (Cabinet Office, Government of Japan) and Collaborative Research Grant from the Institute of Fluid Science, Tohoku University.

## AUTHOR CONTRIBUTIONS

**Conceptualization and supervision:** Saqr

**Software, data curation and formal analysis:** Tupin

**Methodology:** Tupin and Ohta

**Funding acquisition and Resources:** Ohta and Tupin

**Investigation**: Tupin, Saqr and Rashad

**Visualization:** Tupin

**Project administration:** Rashad and Niizuma

**Validation:** Niizuma, Ohta and Tominaga

**Writing:** Saqr, Rashad and Tupin

**Editing and revision:** Niizuma and Ohta

**COMPETING INTERESTS:** Authors declare no competing interests.

**DATA ACCESSIBILITY:** All PIV measurements data presented in the figures as well as raw, un-processed data are available on Dryad.

**Review URL:** https://datadryad.org/stash/share/huHInaJWKqP8QQsLIwBaSi6ZKfSEx-QFnzylhZLuQlI

**Dataset DOI:** https://doi.org/10.5061/dryad.7m0cfxpqq

## REFERENCES:

1 Meng, H., Tutino, V. M., Xiang, J., Siddiqui, A. 2014 High WSS or low WSS? Complex interactions of hemodynamics with intracranial aneurysm initiation, growth, and rupture: toward a unifying hypothesis. *AJNR. American journal of neuroradiology*. **35**, 1254-1262. (10.3174/ajnr.A3558)
2 Yagi, T., Sato, A., Shinke, M., Takahashi, S., Tobe, Y., Takao, H., Murayama, Y., Umezu, M. 2013 Experimental insights into flow impingement in cerebral aneurysm by stereoscopic particle image velocimetry: Transition from a laminar regime. *Journal of the Royal Society Interface*. **10**, (10.1098/rsif.2012.1031)

3 Berg, P., Abdelsamie, A., Yu, H., Janiga, G., Thévenin, D. Year Multi-phase blood flow modeling in intracranial aneurysms considering possible transition to turbulence. ASME 2013 Summer Bioengineering Conference, SBC 2013; 2013; 2013.

4 Valen-Sendstad, K., Mardal, K. A., Steinman, D. A. 2013 High-resolution CFD detects high-frequency velocity fluctuations in bifurcation, but not sidewall, aneurysms. *Journal of biomechanics*. **46**, 402-407. (10.1016/j.jbiomech.2012.10.042)

5 Valen-Sendstad, K., Piccinelli, M., Steinman, D. A. 2014 High-resolution computational fluid dynamics detects flow instabilities in the carotid siphon: implications for aneurysm initiation and rupture? *Journal of biomechanics*. **47**, 3210-3216. (10.1016/j.jbiomech.2014.04.018)

6 Xu, L., Liang, F., Gu, L., Liu, H. 2018 Flow instability detected in ruptured versus unruptured cerebral aneurysms at the internal carotid artery. *Journal of biomechanics*. **72**, 187-199. (10.1016/j.jbiomech.2018.03.014)

7 Jain, K., Roller, S., Mardal, K. A. 2016 Transitional flow in intracranial aneurysms - A space and time refinement study below the Kolmogorov scales using Lattice Boltzmann Method. *Computers and Fluids*. **127**, 36-46. (10.1016/j.compfluid.2015.12.011)

8 Chiu, J. J., Chien, S. 2011 Effects of disturbed flow on vascular endothelium: pathophysiological basis and clinical perspectives. *Physiol Rev*. **91**, 327-387. (10.1152/physrev.00047.2009)

9 Chatterjee, S. 2018 Endothelial Mechanotransduction, Redox Signaling and the Regulation of Vascular Inflammatory Pathways. *Frontiers in physiology*. **9**, 524-524. (10.3389/fphys.2018.00524)

10 Saqr, K. M., Rashad, S., Tupin, S., Niizuma, K., Hassan, T., Tominaga, T., Ohta, M. 2019 What does computational fluid dynamics tell us about intracranial aneurysms? A meta-analysis and critical review. *Journal of Cerebral Blood Flow & Metabolism*. 0271678X19854640.

11 Berg, P., Abdelsamie, A., Janiga, G., Thévenin, D. Year Multi-phase blood flow modelling in an intracranial aneurysm considering possible transition to turbulence. International Symposium on Turbulence and Shear Flow Phenomena, TSFP 2013; 2013; 2013.

12 Xu, L., Zhang, F., Wang, H., Yu, Y. 2012 Contribution of the hemodynamics of A1 dysplasia or hypoplasia to anterior communicating artery aneurysms: A 3-dimensional numerical simulation study. *Journal of Computer Assisted Tomography*. **36**, 421-426. (10.1097/RCT.0b013e3182574dea)

13 Valen-Sendstad, K., Mardal, K. A., Mortensen, M., Reif, B. A. P., Langtangen, H. P. 2011 Direct numerical simulation of transitional flow in a patient-specific intracranial aneurysm. *Journal of biomechanics*. **44**, 2826-2832. (10.1016/j.jbiomech.2011.08.015)

14 Richardson, L. F. 2015 *Weather Prediction by Numerical Process - Scholar's Choice Edition*. Creative Media Partners, LLC.

15 Leslie, D. C., Leslie, D. 1973 *Developments in the Theory of Turbulence*. Clarendon Press Oxford.

16 Birnir, B. 2013 The Kolmogorov–Obukhov statistical theory of turbulence. *Journal of Nonlinear Science*. **23**, 657-688.

17 Kraichnan, R. H. 1967 Inertial ranges in two - dimensional turbulence. *The Physics of Fluids*. **10**, 1417-1423.

18 Sommeria, J. 1986 Experimental study of the two-dimensional inverse energy cascade in a square box. *Journal of fluid mechanics*. **170**, 139-168.

19 Murai, Y., Song, X.-q., Takagi, T., Ishikawa, M.-A., Yamamoto, F., Ohta, J. 2000 Inverse Energy Cascade Structure of Turbulence in a Bubbly Flow: PIV Measurment and Results. *JSME International Journal Series B Fluids and Thermal Engineering*. **43**, 188-196.
20 Lesieur, M., Metais, O. 1996 New trends in large-eddy simulations of turbulence. *Annual review of fluid mechanics*. **28**, 45-82.
21 Van Heijst, G. J. F., Flór, J. B. 1989 Dipole formation and collisions in a stratified fluid. *Nature*. **340**, 212-215. (10.1038/340212a0)
22 Biferale, L., Musacchio, S., Toschi, F. 2012 Inverse energy cascade in three-dimensional isotropic turbulence. *Physical Review Letters*. **108**, (10.1103/PhysRevLett.108.164501)
23 Cho, J. Y. N., Anderson, B. E., Barrick, J. D. W., Lee Thornhill, K. 2001 Aircraft observations of boundary layer turbulence: Intermittency and the cascade of energy and passive scalar variance. *Journal of Geophysical Research Atmospheres*. **106**, 32469-32479. (10.1029/2001JD900079)
24 Young, R. M. B., Read, P. L. 2017 Forward and inverse kinetic energy cascades in Jupiter's turbulent weather layer. *Nature Physics*. **13**, 1135. (10.1038/nphys4227

https://www.nature.com/articles/nphys4227#supplementary-information)
25 Xu, D., Warnecke, S., Song, B., Ma, X., Hof, B. 2017 Transition to turbulence in pulsating pipe flow. *Journal of Fluid Mechanics*. **831**, 418-432. (10.1017/jfm.2017.620)
26 Brindise, M. C., Vlachos, P. P. 2018 Pulsatile pipe flow transition: Flow waveform effects. *Physics of Fluids*. **30**, (10.1063/1.5021472)
27 Çarpinlioğlu, M. O., Özahi, E. 2012 An updated portrait of transition to turbulence in laminar pipe flows with periodic time dependence (a correlation study). *Flow, Turbulence and Combustion*. **89**, 691-711. (10.1007/s10494-012-9420-1)
28 Trip, R., Kuik, D. J., Westerweel, J., Poelma, C. 2012 An experimental study of transitional pulsatile pipe flow. *Physics of Fluids*. **24**, 014103. (10.1063/1.3673611)
29 Stettler, J. C., Hussain, A. K. M. F. 2006 On transition of the pulsatile pipe flow. *Journal of Fluid Mechanics*. **170**, 169-197. (10.1017/S0022112086000848)
30 Holdsworth, D. W., Norley, C. J. D., Frayne, R., Steinman, D. A., Rutt, B. K. 1999 Characterization of common carotid artery blood-flow waveforms in normal human subjects. *Physiological Measurement*. **20**, 219-240. (10.1088/0967-3334/20/3/301)
31 Etminan, N., Rinkel, G. J. 2016 Unruptured intracranial aneurysms: Development, rupture and preventive management. *Nature Reviews Neurology*. **12**, 699-713. (10.1038/nrneurol.2016.150)
32 Valen-Sendstad, K., Mardal, K. A., Mortensen, M., Reif, B. A., Langtangen, H. P. 2011 Direct numerical simulation of transitional flow in a patient-specific intracranial aneurysm. *Journal of biomechanics*. **44**, 2826-2832. (10.1016/j.jbiomech.2011.08.015)
33 Parlea, L., Fahrig, R., Holdsworth, D. W., Lownie, S. P. 1999 An Analysis of the Geometry of Saccular Intracranial Aneurysms. *American Journal of Neuroradiology*. **20**, 1079-1089.
34 M., O., D.A., R. Year Three-dimensional geometry measurements of cerebral aneurysms and vessel sizes for analytic geometry. JSME Fluid Engineering Conference; 2006; 2006. p. 355-356.
35 Kosukegawa, H., Mamada, K., Kuroki, K., Liu, L., Inoue, K., Hayase, T., Ohta, M. 2008 Measurements of Dynamic Viscoelasticity of Poly (vinyl alcohol) Hydrogel for the Development of Blood Vessel Biomodeling. *Journal of Fluid Science and Technology*. **3**, 533-543. (10.1299/jfst.3.533)
36 Ma, D., Dumont, T. M., Kosukegawa, H., Ohta, M., Yang, X., Siddiqui, A. H., Meng, H. 2013 High fidelity virtual stenting (HiFiVS) for intracranial aneurysm flow diversion: in vitro and in silico. *Ann Biomed Eng*. **41**, 2143-2156. (10.1007/s10439-013-0808-4)

37 Ohta, M., Handa, A., Iwata, H., Rüfenacht, D. A., Tsutsumi, S. 2004 Poly-vinyl alcohol hydrogel vascular models for in vitro aneurysm simulations: the key to low friction surfaces. *Technology and Health Care*. **12**, 225-233.
38 Zarrinkoob, L., Ambarki, K., Wåhlin, A., Birgander, R., Eklund, A., Malm, J. 2015 Blood flow distribution in cerebral arteries. *Journal of Cerebral Blood Flow & Metabolism*. **35**, 648-654.
39 Blanco, P. J., Müller, L. O., Spence, J. D. 2017 Blood pressure gradients in cerebral arteries: a clue to pathogenesis of cerebral small vessel disease. *Stroke and vascular neurology*. **2**, 108-117.
40 Cutnell, J. D., Johnson, K. W., Young, D., Stadler, S. 2015 *Physics, 10th Edition*. Wiley.
41 Galduroz, J., Antunes, H., Santos, R. 2007 Gender-and age-related variations in blood viscosity in normal volunteers: a study of the effects of extract of Allium sativum and Ginkgo biloba. *Phytomedicine*. **14**, 447-451.
42 Shida, S., Kosukegawa, H., Ohta, M. 2011 Development of a methodology for adaptation of refractive index under controlling kinematic viscosity for PIV. *Proceedings of the ASME, Colorado USA*.
43 Bouillot, P., Brina, O., Ouared, R., Lovblad, K., Pereira, V. M., Farhat, M. 2014 Multi-time-lag PIV analysis of steady and pulsatile flows in a sidewall aneurysm. *Experiments in fluids*. **55**, 1746.
44 Wieneke, B., Pfeiffer, K. Year Adaptive PIV with variable interrogation window size and shape 15th Int. Symp. on Applications of Laser Techniques to Fluid Mechanics (Lisbon, Portugal, 5–8 July); 2010; 2010.
45 El-Gabry, L. A., Thurman, D. R., Poinsatte, P. E. 2014 Procedure for determining turbulence length scales using hotwire anemometry.
46 Reeves, G. smooth2a function version 1.0. MATLAB Central File Exchange: Mathworks Inc. 2009.
47 Del Álamo, J. C., Jiménez, J. 2009 Estimation of turbulent convection velocities and corrections to Taylor's approximation. *Journal of Fluid Mechanics*. **640**, 5-26. (10.1017/S0022112009991029)
48 Moin, P. 2009 Revisiting Taylor's hypothesis. *Journal of Fluid Mechanics*. **640**, 1-4. (10.1017/S0022112009992126)
49 Kang, W., Lee, S. B., Sung, H. J. 2008 Self-sustained oscillations of turbulent flows over an open cavity. *Experiments in Fluids*. **45**, 693-702. (10.1007/s00348-008-0510-8)
50 Baratchi, S., Khoshmanesh, K., Woodman, O. L., Potocnik, S., Peter, K., McIntyre, P. 2017 Molecular Sensors of Blood Flow in Endothelial Cells. *Trends Mol Med*. **23**, 850-868. (10.1016/j.molmed.2017.07.007)
51 Tarbell, J. M., Simon, S. I., Curry, F. R. 2014 Mechanosensing at the vascular interface. *Annu Rev Biomed Eng*. **16**, 505-532. (10.1146/annurev-bioeng-071813-104908)
52 Yen, W., Cai, B., Yang, J., Zhang, L., Zeng, M., Tarbell, J. M., Fu, B. M. 2015 Endothelial surface glycocalyx can regulate flow-induced nitric oxide production in microvessels in vivo. *PLoS One*. **10**, e0117133. (10.1371/journal.pone.0117133)
53 Shi, Z., Graber, Z. T., Baumgart, T., Stone, H. A., Cohen, A. E. 2018 Cell Membranes Resist Flow. *Cell*. (10.1016/j.cell.2018.09.054)
54 Saqr, K. M., Mansour, O., Tupin, S., Hassan, T., Ohta, M. 2018 Evidence for non-Newtonian behavior of intracranial blood flow from Doppler ultrasonography measurements. *Med Biol Eng Comput*. **In-Press**, (10.1007/s11517-018-1926-9)